\documentclass[aps,prl,twocolumn,showpacs,superscriptaddress,groupedaddress]{revtex4}  
\usepackage{graphicx}  
\usepackage{dcolumn}   
\usepackage{bm}        
\usepackage{amssymb}   

\usepackage{color}
\usepackage{graphicx}
\usepackage{breqn}
\usepackage{dcolumn}
\usepackage{bm}

\usepackage{xspace}

\begin{document}

\widetext

\author{Alexandru B. Georgescu}
\affiliation{Department of Physics}
\affiliation{Center for Research on Interface Structure and Phenomena \\Yale University$,$ New Haven$,$ CT 06520$,$ USA}
\affiliation{Center for Computational Quantum Physics \\The Flatiron Institute$,$ 162 5th Avenue$,$ New York$,$ NY 10010}
\author{Sohrab Ismail-Beigi}
\affiliation{Department of Physics}
\affiliation{Department of Applied Physics}
\affiliation{Department of Mechanical Engineering and Materials Science  \\Yale University$,$ New Haven$,$ CT 06520$,$ USA}
\affiliation{Center for Research on Interface Structure and Phenomena \\Yale University$,$ New Haven$,$ CT 06520$,$ USA}


\title{Surface Piezoelectricity of (0001) Sapphire}

\date{\today}

\begin{abstract}
Interfaces of sapphire are of technological relevance as sapphire is used as a substrate in electronics, lasers, and  Josephson junctions for quantum devices. In addition, its  surface is potentially useful in catalysis.  Using first principles calculations, we show that, unlike bulk sapphire which has inversion symmetry, the  (0001) sapphire surface is piezoelectric. The inherent broken symmetry at the surface leads to a surface dipole and a significant response to imposed strain: the magnitude of the surface piezoelectricity is comparable to that of bulk piezoelectrics.
\end{abstract}

\maketitle

\section{Introduction}
The surface of sapphire ($\alpha$-Al$_2$O$_3$) as well as interfaces between sapphire and other materials have a wide range of electronic or optical applications. As a substrate for the two-dimensional (2D) topological insulator stanene~\citep{Ito2016,Araidai2017} and 2D transition metal dichalcogenides~\citep{2D}, strain effects at the interface are likely to be important. Vibrational modes at the sapphire surface are also likely to couple to vibrational modes in 2D transition metal dichalcogenides which affect their properties, particularly charge density waves and superconductivity~\citep{2D}. In quantum computing devices based on superconductors, alumina plays a critical part in forming the  Josephson junctions~\cite{Josephson}, and interfacial electron-lattice coupling may provide an important loss mechanism.  In fact, the mechanoelectric coupling between surface vibrational modes -- possibly related to the surface piezoelectricity of the surface -- and a piezoelectric has been used to overcome the loss mechanism at the interface between sapphire and aluminum and to design a new type of quantum device \citep{Chu2017}. Finally, another application of sapphire interfaces (in this case, with gold) has been in femtosecond pulse lasers \cite{Spence1991,Han2016}.

Separately and recently, the surface of sapphire has been shown to be useful for catalysis. The hydrophilic nature of aluminum-terminated $\alpha -$Al$_2$O$_3$ (0001) has been exploited to create a scheme for achieving high selectivity in direct methane to methanol~\citep{Latimer2018}, one of the most important goals in present day catalysis. The hydrophilicity of this surface has been traced to lone-pair--surface bonds~\citep{Kakekhani2018}, and the extent of the surface aluminum displacement dictates the interaction of this surface with these technologically relevant molecules.

In ultrahigh vacuum or in an O$_2$ only atmosphere, the aluminum-terminated sapphire (0001) surface is the most stable termination~\citep{Ranea2009,Latimer2018}. The Al ``dangling bond'' at this surface determines many of its properties \cite{Dai2011,Ahn1997,Heffelfinger1997,Batyrev2000,Manassidis1993,Verdozzi1999,Walters2000} through changes in the aluminum position. In the presence of water, the Al-terminated surface becomes hydroxylated~\cite{Nygren1997,Hameka1987,Batyrev2000,Ranea2009}.  Nonetheless, with increased temperature and low H$_2$O pressure, the Al-terminated $\alpha -$Al$_2$O$_3$ (0001) is again stabilized. 

In this work, we use Density Functional Theory (DFT) to investigate the surface polarity, the surface dipole, and how the strain response of the dipole leads to a significant surface piezoelectric effect for both the Al-terminated and hydroxylated surfaces of $\alpha$-Al$_2$O$_3$ (0001). The broad use of Al$_2$O$_3$ as well as recent developments in studying surface piezoelectricity in other materials~\cite{Dai2011} make a study of the piezoelectricity in sapphire an interesting and useful task. In addition, the corundum structure is found in many other materials and the mechanism presented here may be of general interest.  In connection with the above mentioned applications of sapphire, our work can provide a possible microscopic explanation of the efficiency of the above mechanoelectric quantum devices based on surface piezoelectric coupling, highlight new directions for potential modification and control of of polarization-driven surface chemistry~\citep{Kakekhani2018}, and give insight into strain-driven changes in the electronic structure of 2D overlayers due to surface piezoelectric response.

\section{Methodology}
Our calculations are performed using DFT \cite{HK,Kohn1965} with the local density approximation (LDA) \cite{Kohn1965,Perdew1981} using the Quantum Espresso \cite{QE} software package with ultrasoft pseudopotentials \cite{ultrasoft}.  For electronic smearing,  we use a Marzari-Vanderbilt \cite{Marzari1999} smearing width of 0.02 Ry (although all systems turn out to be robust insulators). We use a planewave cutoff of 35 Ryd for the wave functions and 280 Ryd for the electron density.  The k-point sampling uses a $8\times8\times2$ grid for the bulk and a $8\times8\times1$ grid for the slab simulations. We vary the in-plane strain between $\pm 1$\% on our structures to measure the effective out-of-plane piezoelectric response. In the Appendix, we show that this level of strain is well within the linear regime for this material.  Calculations of the bulk dielectric screening and Born effective charges employ the Berry phase method combined with the finite field approach~\cite{Vanderbilt1993,Vanderbilt2002}.

\begin{figure}
\includegraphics[scale=0.23]{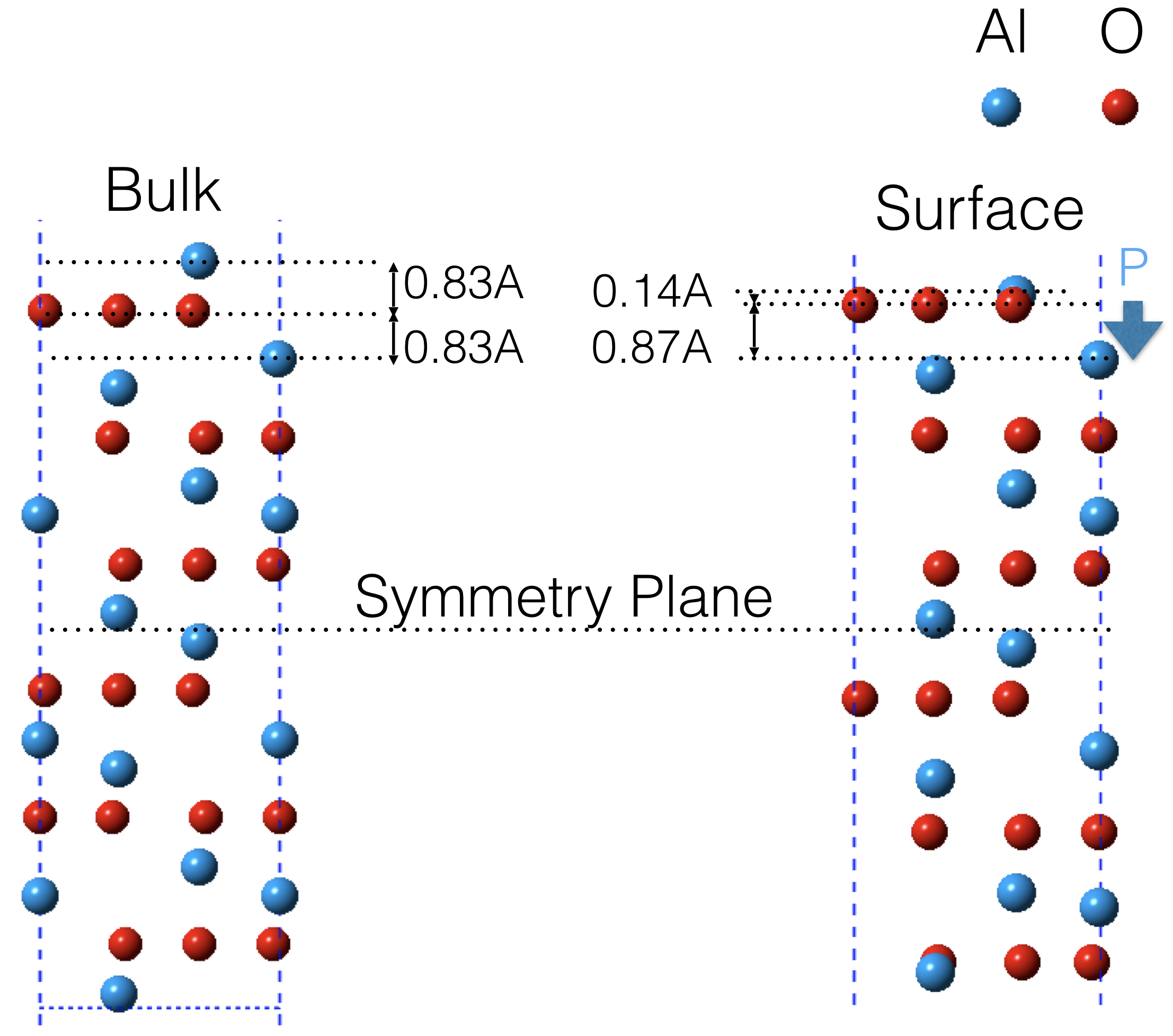}
\caption{\label{fig:BvS} Structures of bulk Al$_2$O$_3$ (left) and the relaxed 30-atom Al-terminated slab structure (right).  The (0001) direction is vertical. Both systems have have a symmetry plane insuring zero net dipole and no long-range polar effects.  The bulk has zero polarization by symmetry while the surface of the slab has strong relaxations leading to a surface dipole. Key surface layer separations are labeled and the direction of surface dipole is indicated by the arrow.}
\end{figure}

We describe bulk Al$_2$O$_3$ using 6 formula units (a 30 atom cell) with a 4.76 \AA\ in-plane ($xy$) lattice constant (19.4 \AA$^2$ in-plane unit cell area) and a $z$ lattice constant of 12.99 \AA: these are theoretical lattice constants that optimize the bulk total energy.  To simulate surfaces, we perform slab simulations using the bulk in-plane lattice constants.  Our slabs have either 30 atoms per unit cell (6 structural units thick) or 60 atoms per unit cell.  We use 30 \AA\  of vacuum added in the $z$ direction to isolate the slabs.  The Al-terminated (0001) surface we use (see Fig.~\ref{fig:BvS}) ensures two identical surfaces, no net dipole for the entire simulation cell, and no stray electric fields in the vacuum region.  While previous studies found that simulating 3 structural units of the surface is sufficient for a good physical picture of the surface of sapphire~\cite{Verdozzi1999}, our thicker slabs permit us to carefully converge the relatively small displacements that penetrate into the sub-surface regions.  Further, we study the hydroxylated (0001) surface for which we follow the same philosophy: starting with the above alumnia slab, we add H$_2$O symmetrically on both sides to generate two identical surfaces.  This results in a 66 atom slab when adding the H$_2$O to the 60 atom alumnia slab.

\section{Results}

\subsection{Al-terminated surface}
  
Our results on the surface structure of Al-terminated (0001) Al$_2$O$_3$ agree with prior literature. We find a bond shortening between the surface Al and the first layer of O~\cite{Dai2011, Ahn1997, Heffelfinger1997, Batyrev2000, Manassidis1993, Verdozzi1999} (see Fig.~\ref{fig:BvS}) that is polar and would, under atmospheric conditions, be hydroxilated due to its polarity \cite{Nygren1997, Hameka1987, Batyrev2000}. Here, our objective is to go beyond this and analyze the dependence of this polarity on applied stress.

In Fig.~\ref{fig:BvS}, we show the structure of the bulk and our 30-atom slab. The bulk structural unit along (0001) is most easily understood as starting with an Al layer, followed by three O atoms in the second layer, and ending with another Al layer. Effectively, this forms two dipoles pointing from the O layer to the Al layers of equal and opposite directions which reflects the inversion symmetry of the bulk. To directly compare to the slab results, we have replicated the basic repeating unit six times when showing the bulk structure in Fig.~\ref{fig:BvS}.  This choice of structural unit is non-polar along (0001), automatically and correctly ensuring no piezoelectric response for the bulk.

\begin{figure}
\includegraphics[scale=0.23]{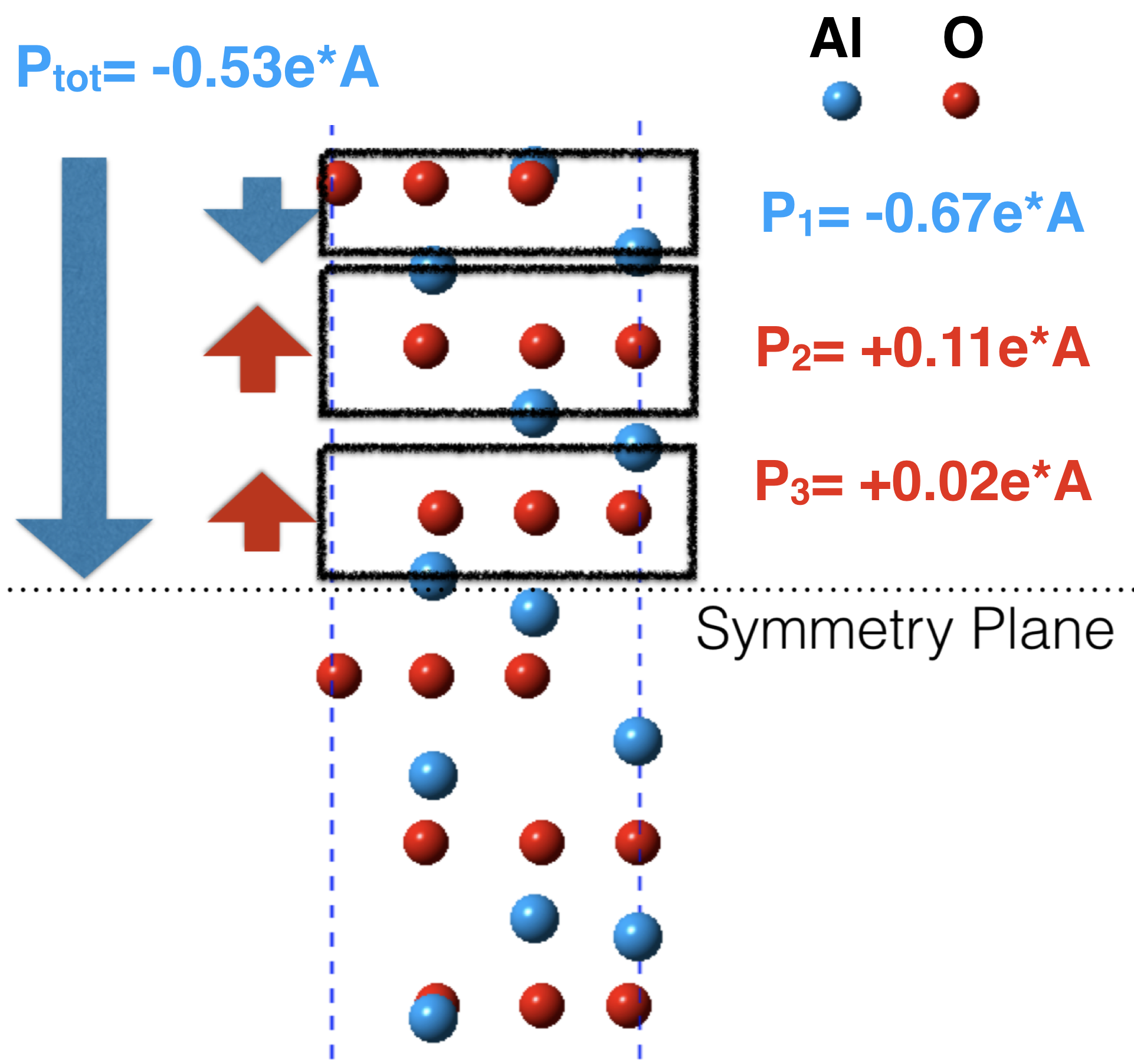}
\caption{Surface dipole moments of the (zero-strain) Al$_2$O$_3$ surface in 30 atom slab with respect to the corresponding bulk configuration. Each five atom structural unit is enclosed by a black rectangle. The first structural unit from the vacuum unit has the largest polarization. Screened bulk Born effective charges are used to compute the vertical components of the dipoles.}
\label{fig:Al2O3} 
\end{figure}

To better understand the nature of the surface relaxations, we divide the top half of the slab into its structural units as shown in Fig.~\ref{fig:Al2O3}.  The relaxations of the top Al layer are very large  (0.69 \AA\ inward motion of the top Al layer as per Fig.~\ref{fig:BvS}):  the surface Al has no O to bond to above it, so it moves into the surface to strengthen its available Al-O bonds. This strong distortion foreshadows a sizable response to perturbations.  The layer-dependent Al-O $z$ separations are displayed in Table \ref{tab:dz}: each structural unit becomes distorted due to its proximity to the surface, but the magnitude of the perturbations decay rapidly going into the material.

We begin with the simplest physical approach that provides  order-of-magnitude estimates of the dipole developed due to surface relaxation and strain.  We sum over the bulk Born effective (dynamical) charges $Z^*$ of the ions multiplied by the ionic displacements to compute dipoles. Bulk Born effective charges are defined under periodic electrostatic boundary conditions while our slab has no such periodicity along $z$, hence the appropriate charges to use are the screened Born charges $\tilde Z^* = Z^*/\epsilon_\infty$ where $\epsilon_\infty$ is the optical (clamped ion) dielectric constant: this reflects the fact that the electrons can, and will, screen the dipole formed by moving an ion in the slab geometry.  We find, using the Berry phase approach, the bulk values $Z^*_{Al}=+2.92$, $Z^*_O=-1.94$ and $\epsilon_\infty=3.20$ which agree well with prior LDA results~\citep{Sapphire}. Note that replacing the $Z^*$ by formal charges Al$^{3+}$ and O$^{2-}$ is a good approximation due to the strongly ionic nature of the material, but neglecting the electronic screening $\epsilon_\infty$ leads to large errors.  Fig.~\ref{fig:Al2O3} and the center column of Table \ref{tab:pvstrain} show dipoles formed per structural unit when using the $\tilde Z^*$ (numerically $+0.913$ for Al$^{3+}$ and $-0.606$ for O$^{2-}$).  The large downward dipole of the topmost (surface) unit is partially canceled by the smaller upward responses of the units below it.  

Next, we use the screened effective charges to address the effect of strain.  The strain can be due to static perturbations (e.g., epitaxy with a substrate or imposed static mechanical stress) or dynamical drive (e.g., acoustic sound waves or dynamic stresses).  We change both in-plane $xy$ lattice parameters biaxially by the same amount and recompute the relaxed structure and layer-by-layer dipoles along $z$ as shown in Table \ref{tab:pvstrain}.
Changing from -1\% (compressive)  to +1\% (tensile) leads to a 1.7\% decrease in the magnitude of the net dipole moment across the entire surface region. If we ascribe this dipole to a region of thickness 2.17 \AA, corresponding to the height of one bulk structural unit in the $z$ direction, then we can estimate a piezoelectric coefficient in units appropriate for comparison to bulk materials.  We find a value of $-3.4\times 10^{-3}$ C/m$^2$ for the change in polarization (dipole per volume) and a piezoelectric coefficient $e_{31}=-0.088$ C/m$^2$ (using Voigt notation and engineering strain convention). The negative sign means the net surface dipole becomes more positive with compressive strain which, as per Table~\ref{tab:dz}, is primarily due to the surface Al cation moving outwards (figuratively, it ``pops out'' due to being ``squeezed'' by its inward moving O neighbors).

To make sure that our results are robust, we also performed calculations using the same methodology using GGA as an exchange-correlation functional, and found that the change is minor (a 3\% increase in the piezoelectric coefficient). Separately, while the specific numerical value of $e_{31}$ depends on the equivalent thickness chosen (2.17 \AA\ above), the rapid decay of dipoles shown in Fig.~\ref{fig:Al2O3} and Table~\ref{tab:pvstrain} means that the appropriate thickness is close to the height of a single bulk structural unit.

\begin{table}
\begin{tabular}{|c |c |c | c | c| } 
 \hline
Str. unit & Atomic planes & +1\% strain &  No strain & -1\% strain  \\ \hline 
 1 & Al-O & 0.112 \AA & 0.143 \AA & 0.173 \AA \\ 
 1 & O-Al & 0.852 \AA & 0.873 \AA & 0.893 \AA \\  \hline
 2 & Al-O & 0.997 \AA & 0.999 \AA & 1.002 \AA \\ 
 2 & O-Al & 0.872 \AA & 0.876 \AA & 0.881 \AA \\  \hline
 3 & Al-O & 0.840 \AA & 0.849 \AA & 0.858\AA \\  
 3 & O-Al & 0.815\AA & 0.828 \AA & 0.841 \AA \\  \hline
  4 & Al-O & 0.815 \AA & 0.829 \AA & 0.842 \AA \\  
  4 & O-Al & 0.824 \AA & 0.836 \AA & 0.848 \AA \\  \hline
  5 & Al-O & 0.826 \AA & 0.837 \AA & 0.849 \AA \\ 
 5 & O-Al & 0.825 \AA & 0.837 \AA & 0.849 \AA \\  \hline
  6 & Al-O & 0.825 \AA & 0.837 \AA & 0.849 \AA \\ 
 6 & O-Al & 0.825 \AA & 0.837 \AA & 0.849 \AA \\ 
 \hline
\end{tabular}
\caption{Separation along $z$ of Al planes and their neighboring O planes within the same structural units of the 60 atom slab. Structural units are counted as starting at the surface and going into the slab. Negative strain is compressive and positive strain is tensile.}
\label{tab:dz}
\end{table}

\begin{table}
\begin{tabular}{ |c | r | r | r | } 
 \hline
 Structural Unit & +1\%  Strain & 0 Strain & -1\% Strain  \\ \hline 
   1 &  -0.675 e\AA    &  -0.666 e\AA  &  -0.657 e\AA \\  \hline
   2 & +0.114 e\AA      & +0.112 e\AA   &   +0.110 e\AA  \\ \hline 
   3 & +0.023 e\AA      & +0.019 e\AA   &  +0.015 e\AA \\ \hline  
   4 & -0.008 e\AA     & -0.007 e\AA   &  -0.005 e\AA \\ \hline 
   5 & +0.0005 e\AA     & +0.0005 e\AA   &  +0.0004 e\AA \\ \hline  
   6 & +0.0002 e\AA      & +0.0002 e\AA   &  +0.0002 e\AA \\ \hline  
   Sum & -0.545 e\AA & -0.541 e\AA   &  -0.536 e\AA \\ \hline
\end{tabular}
\caption{Strain-dependent (0001) dipoles (relative to bulk) for each structural unit of the (0001) Al$_2$O$_3$ slab with 60 atoms as well as their sum. Negative strain is compressive and positive strain is tensile. Biaxial strain is imposed in the $xy$ plane. Screened bulk Born effective charges are used to compute the dipoles. }
\label{tab:pvstrain}
\end{table}
  
\begin{figure}
\includegraphics[scale=0.23]{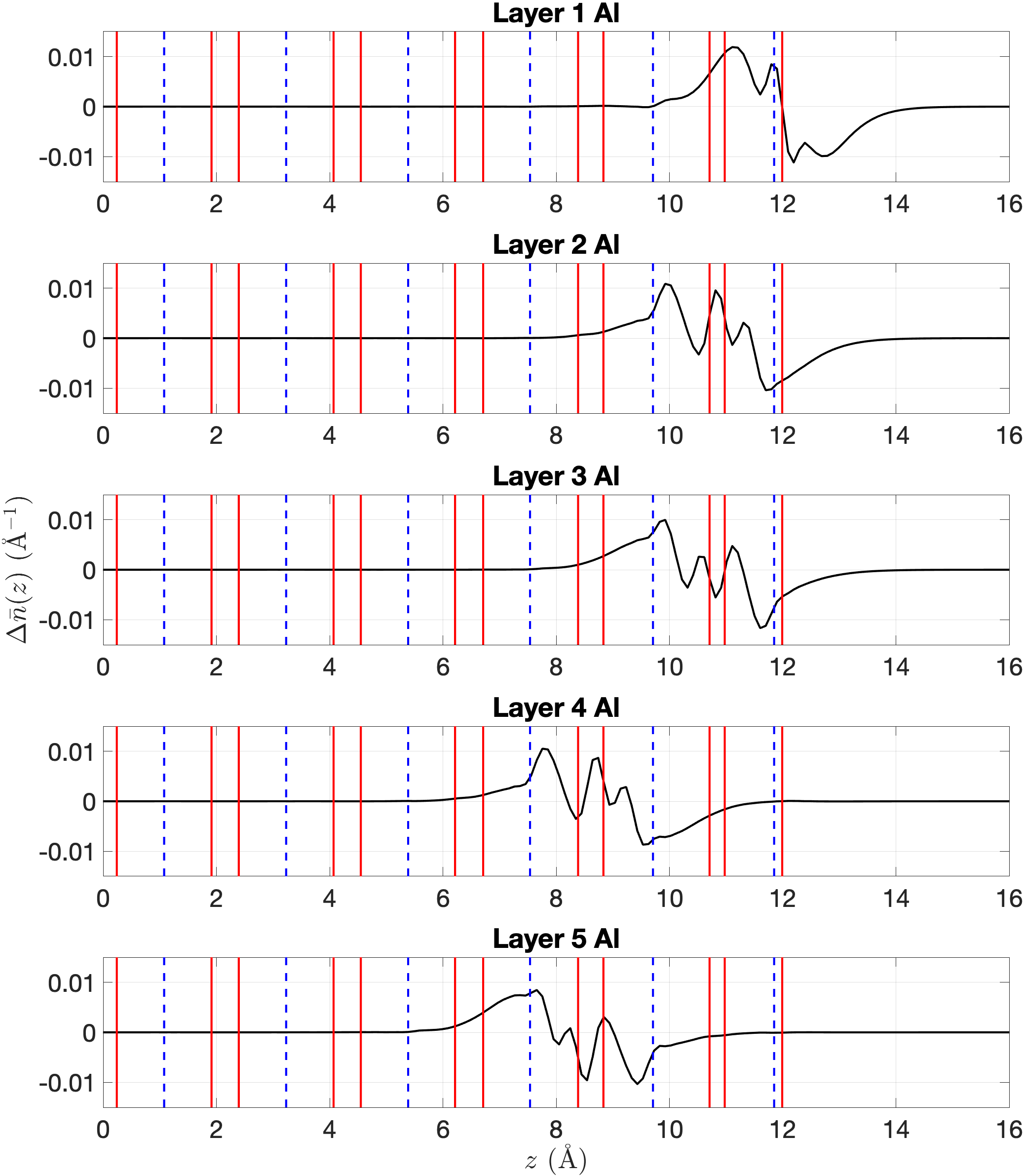}
\caption{Plane-averaged change of electron density due to displacing Al ions by $\delta z=+0.01$ \AA\ near the surface of (0001) Al$_2$O$_3$ in each two-dimensional atomic plane.  Layer 1 is the surface Al layer and increasing layer numbers move into the material.  Black curves show the change of electron density (averaged in the $xy$ plane).  Vertical solid red lines indicate the $z$ coordinates of Al planes while blue dashed vertical lines indicate $z$ positions of O planes.  The origin is the center of the slab.  The electron density changes (combined with the motion of the ionic core and nucleus) correspond to effective charges of $+0.83$, $+0.77$, $+0.81$, $+0.89$, and $+0.94$ when going from layer 1 to layer 5 which are to be compared to the  screened bulk value of $+0.913$.}
\label{fig:edensshiftal} 
\end{figure}

\begin{figure}
\includegraphics[scale=0.25]{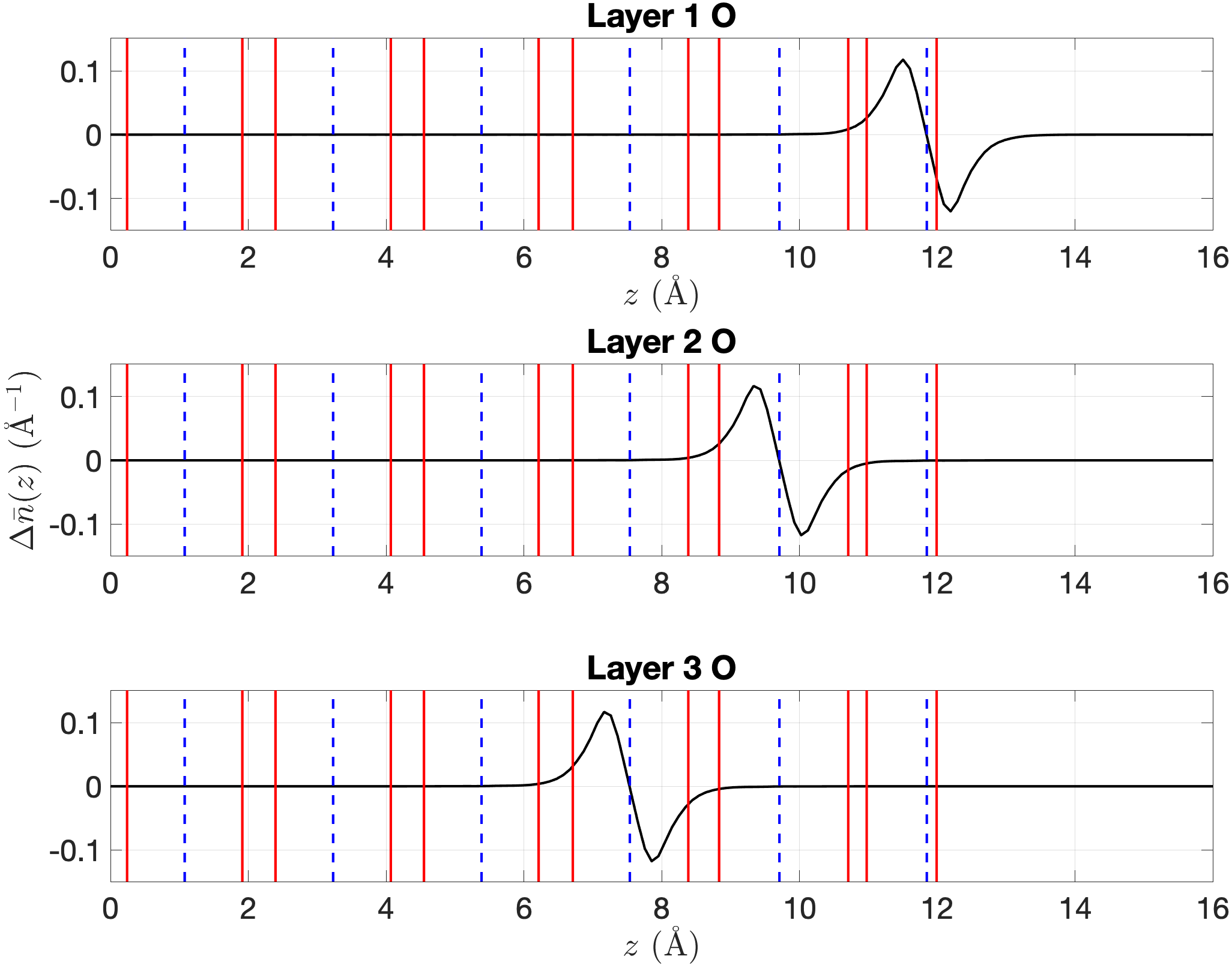}
\caption{Plane-averaged change of electron density due to displacing O ions by $\delta z=+0.01$ \AA\ near the surface of (0001) Al$_2$O$_3$ in each two-dimensional atomic plane.  Same nomenclature as Fig.~\ref{fig:edensshiftal}.  The electron density changes (combined with the motion of the ionic core and nucleus) correspond to effective charges of $-0.50$, $-0.61$, and $-0.61$ when going from layer 1 to layer 3 which are to be compared to the screened bulk value of $-0.606$.}
\label{fig:edensshifto} 
\end{figure}

The above analysis provides a physical understanding of the surface response and semi-quantitative results.  A more quantitative calculation based on Born effective charges requires incorporating changes to the values of the $\tilde Z^*$ for the ions near the surface via cumbersome explicit calculations.  We find that the $\tilde Z^*$ of the ions at or near the surface are reduced from bulk values.  For the Al cations with their larger valence atomic orbitals, the Al in the first 3 layers have their $\tilde Z^*$ reduced by $\sim$10-15\% (deeper layers are within 3\% of bulk values).  For the O anions with their smaller valence atomic orbitals, the O in the surface layer suffers an 18\% reduction of $\tilde Z^*$ while the deeper layers have values within 1\% of the bulk value.  Figures \ref{fig:edensshiftal} and \ref{fig:edensshifto} present these trends graphically by displaying the changes of electron density when individual ionic layers are moved: the shift of electron density is much wider when moving Al cations compared to O anions, and the convergence to the bulk distribution is much faster for O than for Al (also shown numerically  in the  captions).

\begin{figure}
\includegraphics[width=3in]{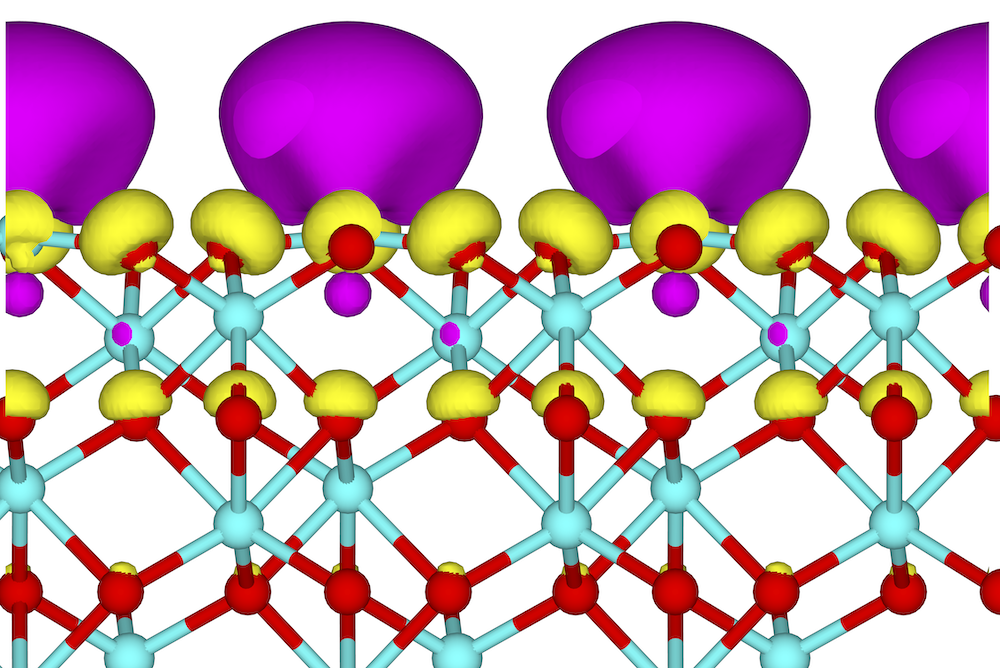}
\caption{Isosurface plot of the conduction band edge wave function of the unperturbed (relaxed) Al$_2$O$_3$ surface.  Purple indicates a positive value and yellow a negative value of the same magnitude.  The state is strongly localized on the surface Al and shaped much like ``dangling'' orbitals pointing out of the surface.}
\label{fig:cbm3dplot} 
\end{figure}

\begin{figure}
\includegraphics[width=3in]{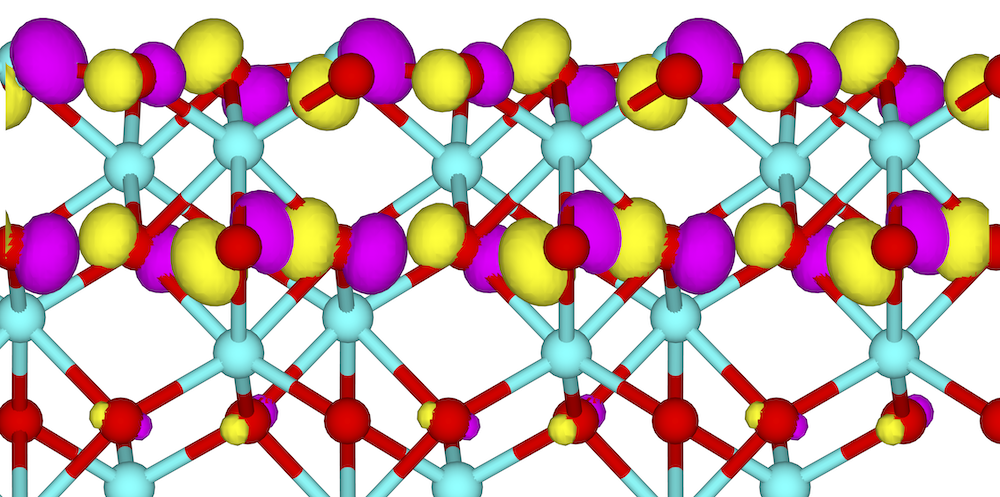}
\caption{Isosurface plot of the valence band edge wave function of the unperturbed (relaxed) Al$_2$O$_3$ surface.  Purple indicates a positive value and yellow a negative value of the same magnitude.  The state is essentially of pure O 2p character and is localized primarily on the surface and subsurface oxygen layers.}
\label{fig:vbm3dplot} 
\end{figure}

Qualitatively, the fact that the surface region shows reduced $\tilde Z^*$ values is sensible since the band gap near the surface is smaller that the bulk and broken/dangling bonds at the surface are more polarizable.  To understand this point in more detail, Figures \ref{fig:cbm3dplot} and \ref{fig:vbm3dplot} display the nature of the conduction and valence band edges of the unstrained (relaxed) alumnia slab.  The conduction band edge state is localized almost exclusively on the surface Al layer with a  directional ``dangling'' out-of-surface shape, while the valence band edge state is a superposition of O 2p states that are localized on the surface layer and the subsurface layer.  Hence, the band edge states are indeed localized on the surface.  Furthermore we expect the conduction band edge state to be much more sensitive to the existence of the surface than the valence band edge.  We corroborate this sensitivity by computing the change of energy of the band edges upon formation of the surface. The LDA band gap of the slab is 4.8 eV which is 1.6 eV lower than the LDA bulk band gap of 6.4 eV.  By using the minimum energy of the localized O 2s band as an energy reference, we find that the band gap reduction is essentially due to the conduction band edge dropping in energy by $\approx 1.5$ eV (while the valence band edges rises by a small amount of $\approx 0.04$ eV).  The lowering of the conduction band edge energy is sensible since the surface Al cations have reduced coordination: the removal of negatively charged O cation nearest neighbors reduces the electrostatic (Madelung) repulsive potential felt by electrons on the Al sites; in parallel, removal of Al-O bonds for the surface Al cations reduces the anti-bonding character, and thus lowers the energy, of the Al-dominated conduction band states.

However, instead of pursuing a cumbersome approach based on screened effective charges, the slab geometry with vacuum permits us to compute the change of surface dipole in two independent and straightforward ways using the electron density itself.
First, the {\it ab initio} calculations provide us with the total charge density $\rho$, and since we have vacuum in our unit cell where $\rho$ essentially drops to zero, computation of the total dipole $p_z=\int d^3r\,z\,\rho$ in the surface region of the slab is straightforward.  To do this, the integral is replaced by a discrete sum over the real-space grid used in the calculation, and to find consistent results, we decompose the integral over the surface layers of one side of the slab into a sum of dipolar contributions from contiguous material units where each unit has an pseudo-electron density integrating 24 pseudo-valence electrons (the correct number for one Al$_2$O$_3$ unit); after the first few surface units, the dipole for each further unit is essentially zero (as it necessary from the symmetry of bulk sapphire) and we terminate the integral.  By computing the change of this dipole between the 1\% tensile and 1\% compressive strain states, we obtain a change of polarization equal to $-6.1\times 10^{-3}$ C/m$^2$ and $e_{31}=-0.16$ C/m$^2$. 

\begin{figure}
\includegraphics[width=3in]{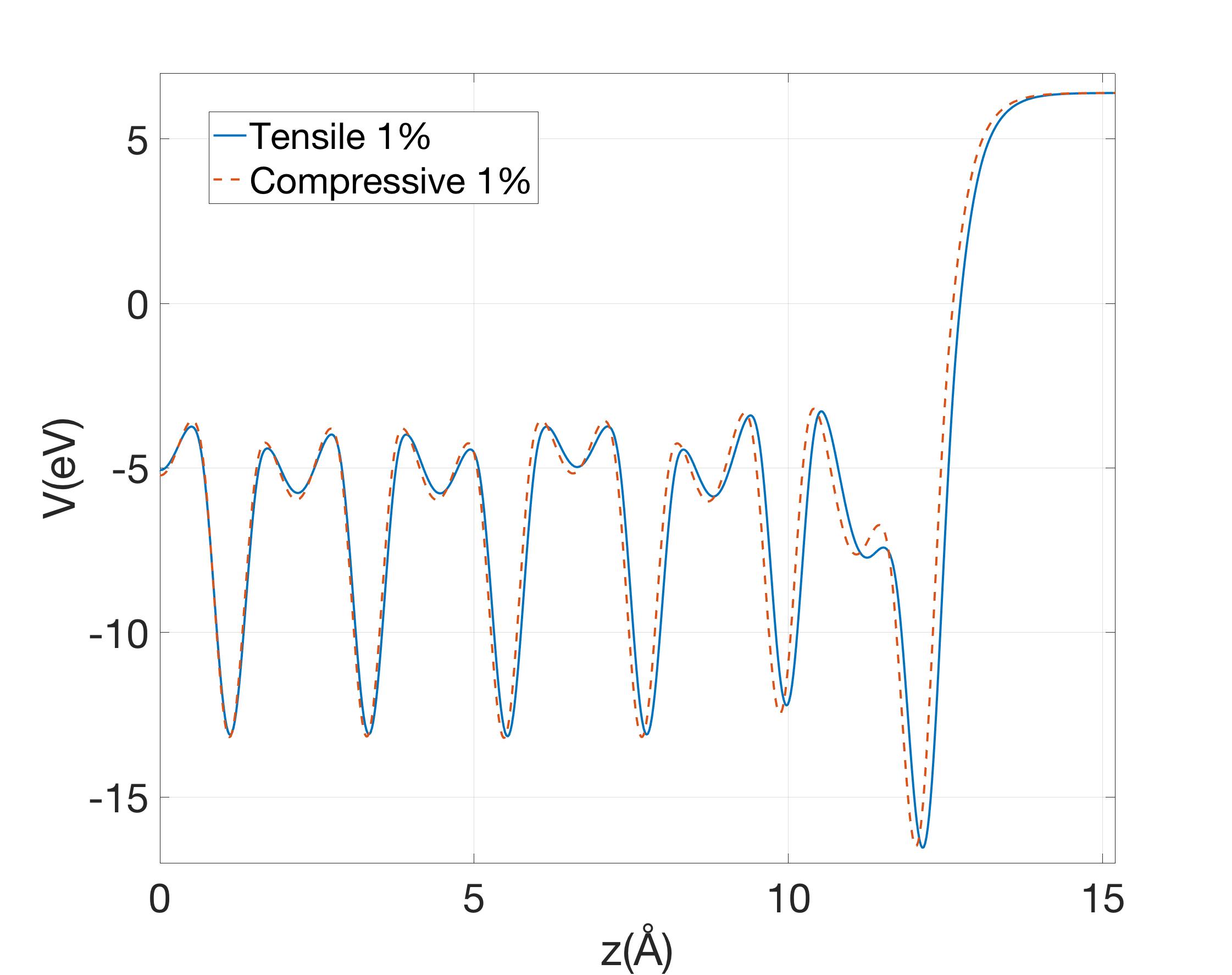}
\caption{Plane-averaged electrostatic potential along $z$ for 1\% tensile and compressive strain, with $z=0$ defined to be the center of the slab, and the surface is at $z\approx13$ \AA.  The potential in the vacuum is shifted so that it has the same value for the two strain states. The change in electrostatic potential described in the main text is obtained by taking the difference of potentials at $z=0$.}
\label{fig:potplot} 
\end{figure}

Second, using basic electrostatics, in a parallel plate geometry such as our slab calculation, a dipole along $z$ will create a potential drop across it, and if we examine the change of potential versus strain, we can back out the change of dipole.  Hence, we align the central symmetry plane of the slabs in all our calculations and compute the electrostatic potential for each simulation averaged in the $xy$ plane.  We then take the difference between the potential on the symmetry plane at the center of our slab and deep in the vacuum where the potential is constant. Figure~\ref{fig:potplot} shows the two potential profiles.  The strain dependence of this potential difference is then computed. Between the $\pm1\%$ strain states examined, we find a net potential difference of $\approx0.16$ V.  A parallel plate model together with an assumed thickness of 2.17 \AA\  gives us a net change of polarization of $-6.5 \times 10^{-3}$ C/m$^2$.  This translates into $e_{31}=-0.17$ C/m$^2$. 

\subsection{Hydroxylated surface}
\begin{figure}
\includegraphics[width=3in]{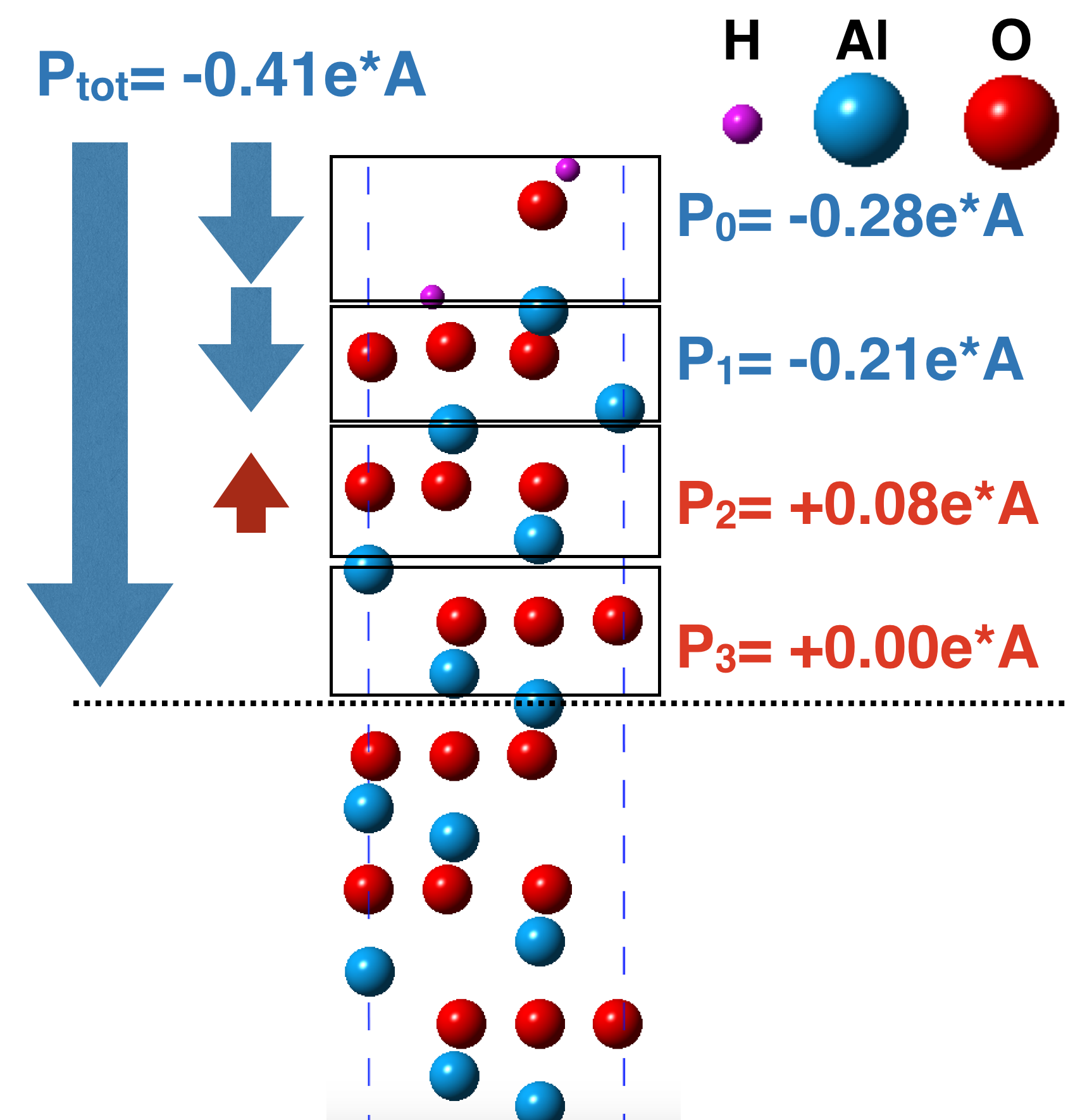}
\caption{Surface structure and dipole moments of the zero-strain and relaxed hydroxylated Al$_2$O$_3$ surface in the 30 atom slab with respect to the corresponding bulk configuration. The surface water molecule followed by the first three unit cells of Al$_2$O$_3$ are framed with black triangles. The water molecule and the first sapphire unit cell have the largest dipole moment. Screened bulk Born effective charges are used. }
\label{fig:h2osurf} 
\end{figure}

Our final set of results concern the effect of changing the surface termination.  To examine this effect, we consider the hydroxylated Al-terminated (0001) sapphire surface with an H$_2$O coverage of 1 H$_2$O molecule per primitive surface unit cell.  This surface has been studied by both theory and experiment \cite{Nygren1997,odziana2003,Eng2000}, and it is known that the H$_2$O molecule dissociates into a hydroxyl radical (OH)$^-$ that bonds via its O to the surface Al cation, and a hydrogen H$^+$ that bonds to an oxygen anion in the layer immediately below the terminating Al layer.  The structure of the surface region is shown in Figure~\ref{fig:h2osurf} along with the dipoles of the various structural units.  Table~\ref{tab:dzh2oterm} shows the vertical separations of cations and anions for the hydroxylated surface (for all three strain values), and Table~\ref{tab:pvstrainh2o} shows the dipoles of the structural units at and near the surface.

\begin{table}
\begin{tabular}{| c | c | c | c | c | } 
 \hline
Struct. Unit & Atomic Plane & +1\% Strain &  0 Strain & -1\% Strain  \\ \hline 
0 & H-O & 0.578 \AA & 0.583 \AA & 0.586 \AA \\
0 & O-H & 1.435 \AA & 1.477 \AA & 1.517 \AA \\ \hline
 1 & Al-O & 0.650 \AA & 0.669 \AA & 0.687 \AA \\ 
 1 & O-Al & 0.879 \AA & 0.890 \AA & 0.901 \AA \\  \hline
 2 & Al-O & 0.918 \AA & 0.921 \AA & 0.925 \AA \\ 
 2 & O-Al & 0.827 \AA & 0.838 \AA & 0.849 \AA \\  \hline
 3 & Al-O & 0.815 \AA & 0.827 \AA & 0.841 \AA \\  
 3 & O-Al & 0.820 \AA & 0.832 \AA & 0.845 \AA \\  \hline
 4 & Al-O & 0.823 \AA & 0.834 \AA & 0.847 \AA \\  
 4 & O-Al & 0.825 \AA & 0.837 \AA & 0.849 \AA \\  \hline
 5 & Al-O & 0.825 \AA & 0.837 \AA & 0.849 \AA \\  
 5 & O-Al & 0.825 \AA & 0.836 \AA & 0.848 \AA \\  \hline
 \hline
\end{tabular}
\caption{Separation along $z$ of cations and anions for the
hydroxylated surface for each structural unit of the 66 atom slab: the surface H$_2$O is structural unit zero, the remaining structual units are the same as for the Al-terminated surface.  Negative strain is compressive and positive strain is tensile.  The values listed in the table for structural units 1-5 are averages over the three inequivalent oxygen anions in each structural unit.}
\label{tab:dzh2oterm}
\end{table}

\begin{table}
\begin{tabular}{ |c | r | r | r | } 
 \hline
 Structural Unit & +1\%  Strain & 0 Strain & -1\% Strain  \\ \hline 
 0 & -0.268 e\AA    & -0.272 e\AA  &  -0.283 e\AA \\  \hline
 1 & -0.209 e\AA    & -0.202 e\AA  &  -0.195 e\AA \\  \hline
 2 & +0.083 e\AA    & +0.076 e\AA  &  +0.070 e\AA  \\ \hline 
 3 & -0.004 e\AA    & -0.004 e\AA  &  -0.004 e\AA \\ \hline  
 4 & -0.002 e\AA    & -0.002 e\AA  &  -0.002 e\AA \\ \hline 
 5 & +0.001 e\AA    & +0.001 e\AA  &  +0.000 e\AA \\ \hline  
Sum & -0.394 e\AA   & -0.403 e\AA  &  -0.413 e\AA \\ \hline
\end{tabular}
\caption{Strain-dependent (0001) dipoles (relative to bulk) for each structural unit of the hydroxylated surface of the 66 atom slab. Negative strain is compressive and positive strain is tensile. Biaxial strain is imposed in the $xy$ plane. Screened bulk Born effective charges are used. }
\label{tab:pvstrainh2o}
\end{table}

We begin with the observation that, due to being capped by the (OH)$^-$, the surface Al is pulled out of the surface compared to the clean Al-terminated case. Next, despite the higher position of the Al cation, the net surface dipole for the hydroxylated surface points inward just as for the Al-terminated case.  But this time, the net inward dipole is dominated by the dipole contribution of the dissociated H$_2$O unit: as is visible from Figure~\ref{fig:h2osurf}, the H$^+$ sits well below the (OH)$^-$ making for a large net negative dipole contribution.  

However, although both surface terminations have a net negative dipole, their strain responses have opposite signs: unlike the Al-terminated case, compressive strain on the hydroxylated surface makes for a more negative surface dipole.  Examining the individual structural unit contributions in Table~\ref{tab:pvstrainh2o}, we see that the alumina subsystem behaves similarly to the clean Al-terminated case: its net dipole becomes more positive with compressive strain. However, the response of the surface H$_2$O unit is negative and larger in magnitude leading to a net negative response.  Analysis of the layer-by-layer vertical separations and their strain response paints the following picture: the surface Al moves up with compressive strain, and this in turn moves its capping (OH)$^-$ rigidly upwards (``riding'' behavior); however, the H$^+$ does not move much since it is bound to the next subsurface O layer.  The net effect is to increase the vertical separation between H$^+$ and (OH)$^-$ and create a more negative surface dipole.

For a numerical estimate of the surface piezoelectric response, we use screened Born effective charges. For Al and O, we use the same values as we did for the Al-terminated surface, while the screened Born effective charge of H is fixed by the neutrality of an H$_2$O molecule (i.e., $2\tilde Z^*_H+\tilde Z^*_O=0$).  These effective charges are used to compute the dipoles shown in Figure~\ref{fig:h2osurf} and Table~\ref{tab:pvstrainh2o}.  This approach leads to an estimated value of $e_{31}=+0.18$ C/m$^2$.  For a more numerically accurate value than can be provided by bulk screened Born effective charges, we apply the dipole integration method described above to the hydroxylated case.  The result is $e_{31}=+0.16$ C/m$^2$ for the hydroxylated surface.

\section{Discussion}

The values find for the surface piezoelectric response coefficient of (0001) sapphire are $|e_{31}|\sim 0.1$.  This  is comparable with workhorse bulk piezoelectric materials such as ZnO with $e_{31}=0.53$ C/m$^2$, LiNbO$_3$ with $e_{31}=0.37$ C/m$^2$, AlN with $e_{31}=0.58$ C/m$^2$, or BaTiO$_3$ with $e_{31}=2.16$ C/m$^2$~\citep{matproj,DeJong2015}.  Hence, the piezoelectricity of the (0001) surface Al$_2$O$_3$ is comparable in strength to that of a unit cell of typical bulk crystalline piezoelectric materials. 

One may also compare our results to appropriately defined surface piezoelectric coefficients from first principles theory \citep{Shen2010,Dai2011}.  For a bulk piezoelectric material such as ZnO, its (0001) surface has a surface piezoelectric constant of $1.0\times10^{-10}$ C/m while the surfaces of non-piezoelectric phases have surface piezoelectric constants that are about ten times smaller \cite{Dai2011}.  For comparison, our Al-terminated Al$_2$O$_3$ (0001) surface has a constant of $3.7\times10^{-11}$ C/m, which is again comparable. 

We note that if one seeks a simple order of magnitude estimate of $|e_{31}|$, one can use formal charges to compute the surface dipole of the relaxed zero-strain surface (a refined estimate screens the surface dipole by dividing by the optical dielectric constant $\epsilon_\infty$).  Then, as Tables~\ref{tab:pvstrain} and \ref{tab:pvstrainh2o} demonstrate, the change of dipole is roughly equal to the strain imposed times the dipole itself, yielding an easy estimate of the dipole response to strain.  Hence, as long as the surface structure of an ionic material is known (through experiment and/or theory), one can quickly provide a ballpark estimate of the strength of the surface piezoelectric response.

Concerning surface termination effects in experimental situations, while the clean Al-terminated and the hydroxylated surfaces we modeled have oppositely signed $e_{31}$ coefficients, the hydroxylated case examined corresponds to the maximum surface coverage for chemisorbed H$_2$O.  For lower concentrations of chemisorbed H$_2$O, we expect $e_{31}$ to have the same sign and magnitude as that of the clean Al-terminated surface.  In addition, for electronic or quantum device applications, any H$_2$O on the surface can easily be driven off using modest temperature increases and vacuum conditions leading to the clean Al-terminated case: we believe the Al-terminated surface to be relevant for such device applications.  For catalytic applications in hydrated or aqueous conditions, the hydroxylated surface response should be the dominant observed behavior.  For either case, our simulations describe ordered crystalline surfaces: understanding the potential effect of surface disorder is more challenging from first principles and beyond the scope of this work.

\section{Conclusion}
Our calculations predict that the (0001) surface of sapphire (Al$_2$O$_3$) has a strong structural response to strain that leads to surface piezoelectricity. Given the wide use of sapphire substrates in many scientific and technological fields, this result is important for understanding and engineering these systems. Furthermore, our work illustrates a more general situation where the surface of a centrosymmetric material can develop significant piezoelectric response due to the broken translational symmetry, and our approach, analysis, and rules of thumb can be applied in a variety of other such materials.

\section{Acknowledgements}
We would like to thank Robert Schoelkopf and Peter T. Rakich for highlighting the importance of sapphire as a key platform for quantum devices. We thank Arvin Kakekhani, Frederick Walker and Claudia Lau for helpful conversations during the writing of this work. The work done at Yale was supported by the National Science Foundation via Grant MRSEC DMR 1119826. We thank the Yale Center for Research Computing for guidance and use of the research computing infrastructure.  This work also used the Extreme Science and Engineering Discovery Environment (XSEDE), which is supported by National Science Foundation grant number ACI-1548562. This work used the Extreme Science and Engineering Discovery Environment (XSEDE) Comet Compute time through allocation MCA08X007, as well as the local Flatiron Institute Iron Cluster. The Flatiron Institute is supported by the Simons Foundation.

\section{Appendix}
\begin{figure}
\includegraphics[scale=0.115]{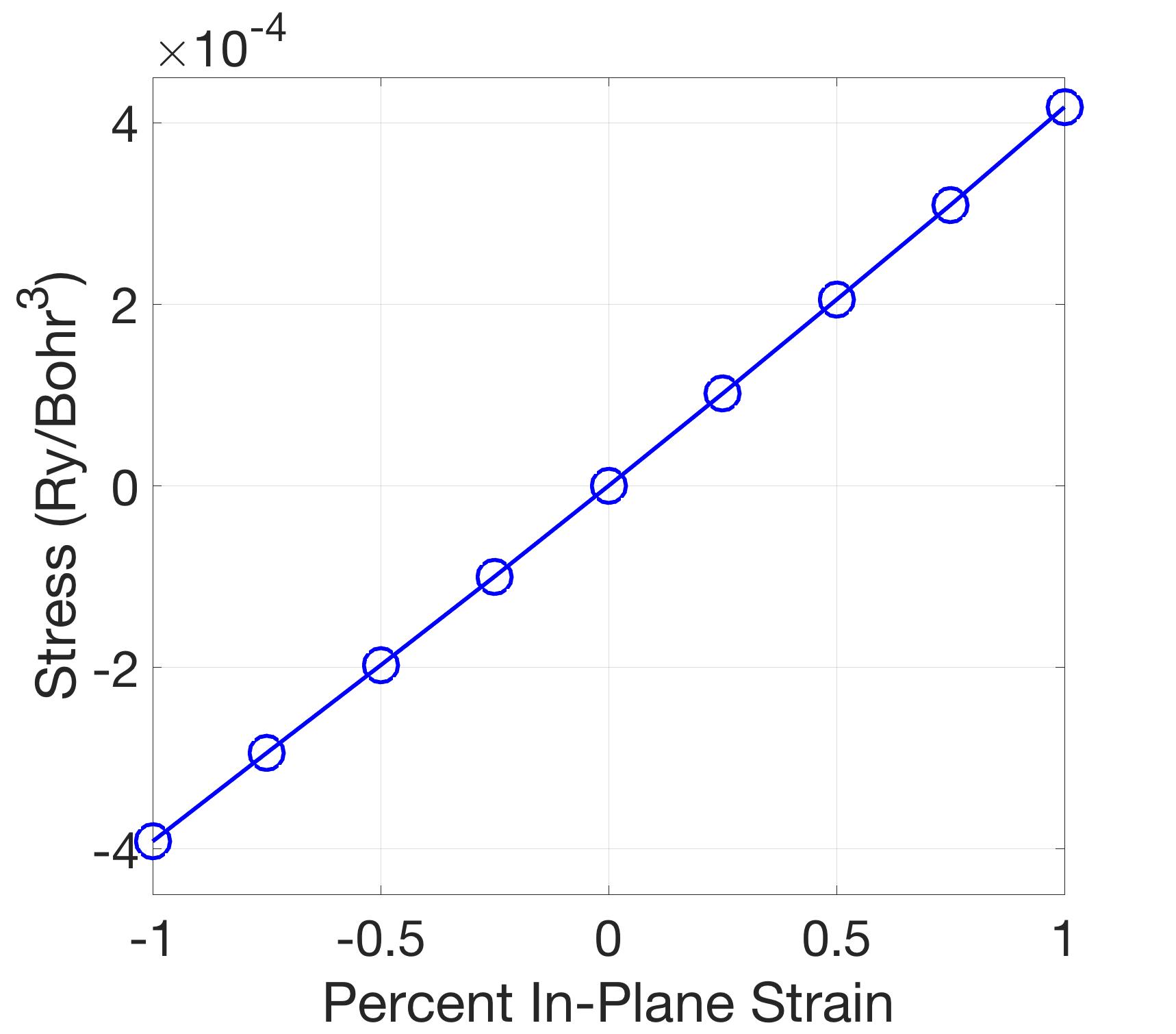}
\caption{Stress versus strain for bulk sapphire based on LDA calculations: compressive strain is positive, and tensile is negative.}
\label{fig:Al2O3strain} 
\end{figure}
To check that our results do not strongly depend on the exchange-correlation functional we pick, we also performed them within the PBE version of the GGA exchange correlation functional \citep{PBE}, and compared the change in dipole with respect to the same change in strain ($\pm 1 \% $) and obtained  minor increase in the estimated dipole change of 3 percent.

Furthermore, in order to check that our calculations are performed in the linear regime of the strain for bulk sapphire, we performed multiple bulk LDA calculations and plotted the stress as a function of strain in Figure~\ref{fig:Al2O3strain}.  The figure shows that the results in the main text are safely in the linear (small strain) regime.

\bibliography{main.bbl}

\end{document}